\documentclass[a4paper,12pt,preprint,amsmath,showpacs,nofootinbib,superscriptaddress,tightenlines]{revtex4-1}
\usepackage{color}
\usepackage{slashed}
\usepackage{graphicx}   
\usepackage[lofdepth,lotdepth]{subfig}
\usepackage{psfrag}
\usepackage{eufrak}
\usepackage{subfig}
\usepackage{tabularx}

\newcommand{\dd}{\mathrm{d}}

\newcommand{\als}{\alpha_s}
\newcommand{\e}{\epsilon}

\newcommand{\nn}{\nonumber}

\newcommand{\Ord}{\mathcal{O}}

\newcommand{\breakl}{\right.\nn\\ &&\left. }
\newcommand{\breakll}{\right.\nn\right.\\ &&\left.\left. }
\def\n3lo{N$^3$LO}
\newcommand{\ca}{C_A}
\newcommand{\crr}{C_i}
\newcommand{\cf}{C_F}
\newcommand{\nf}{n_f}
\newcommand{\cd}{\mathcal{D}}

\allowdisplaybreaks

\begin{document}

\preprint{MITP/14-029, SLAC-PUB-15946}

\title{N$^3$LO Higgs and Drell-Yan production at threshold:\\ the one-loop two-emission contribution}

\author{Ye Li}
\email{yli@slac.stanford.edu}
\affiliation{SLAC National Accelerator Laboratory, Stanford University, Stanford, CA 94309, USA}
\author{Andreas von Manteuffel}
\email{manteuffel@uni-mainz.de}
\author{Robert M. Schabinger}
\email{rschabin@uni-mainz.de}
\affiliation{The PRISMA Cluster of Excellence and
Mainz Institute of Theoretical Physics,
Johannes Gutenberg Universit\"at,
55099 Mainz, Deutschland}
\author{Hua Xing Zhu}
\email{hxzhu@slac.stanford.edu}
\affiliation{SLAC National Accelerator Laboratory, Stanford University, Stanford, CA 94309, USA}

\begin{abstract}
\noindent
In this paper, we study phenomenologically interesting soft radiation distributions in massless QCD. Specifically, we consider the
emission of two soft partons off of a pair of light-like Wilson lines, in either the fundamental or the adjoint representation, at next-to-leading order. Our results are an essential component of the next-to-next-to-next-to-leading 
order threshold corrections to both Higgs boson production in the gluon fusion channel and Drell-Yan lepton production. Our calculations are consistent with the recently published results for Higgs boson production. As a
non-trivial cross-check on our analysis, we rederive a recent prediction for the Drell-Yan threshold cross section using a completely different strategy. 
Our results are compact, valid to all orders in the dimensional regularization parameter, and expressed in terms of pure functions.
\end{abstract}

\maketitle

The total cross section for Higgs boson production in the gluon fusion channel and the total cross section for Drell-Yan lepton production
are two of the most important Large Hadron Collider-relevant, infrared-collinear~(IRC) finite observables in QCD.
Due to their relatively simple kinematics, gluon fusion Higgs and Drell-Yan lepton production are benchmark processes
for the application of perturbative QCD techniques. A great deal of progress along these lines has been made over the past two
decades and, as a consequence, total cross sections for these processes are known to
next-to-next-to-leading order~(NNLO)~\cite{NUPHA.B157.461,NUPHA.B359.283,NUPHA.B359.343,PRLTA.70.1372,hep-ph/9504378,hep-ph/9511344,hep-ph/9705337,hep-ph/0102227,hep-ph/0201206,hep-ph/0207004,hep-ph/0302135,hep-ph/0509189}.
The calculation of these IRC finite observables to one order higher in the strong coupling constant is highly desirable due to the large $K$-factors and theoretical uncertainties inherent in the calculation
of such hadron collider observables~\cite{NUPHA.B359.283,PRLTA.70.1372,hep-ph/9504378,hep-ph/9511344,hep-ph/9705337,hep-ph/0102227,hep-ph/0201206,hep-ph/0207004,hep-ph/0302135}.
Unfortunately, this task is also highly non-trivial;
at next-to-next-to-next-to-leading order~(\n3lo), the current state-of-the-art, one must compute three-loop virtual corrections, single-emission, two-loop real-virtual corrections, double-emission, one-loop real-virtual corrections,
triple-emission real corrections, and collinear counterterms for the parton distribution functions. 
All of these ingredients have severe infrared and/or collinear divergences which complicate their calculation and IRC finiteness is only achieved once all contributions have been appropriately combined.

While full fixed-order calculations at \n3lo remain a challenge, significant progress
has been made over the last few years. For gluon fusion Higgs production, the Wilson coefficients obtained by
integrating out top quark loops to construct an effective $g g h$ Lagrangian are known to
three loops~\cite{NUPHA.B510.61,JHEPA.0601.051,NUPHA.B744.121} and the
relevant ultraviolet renormalization constants have been worked out in
Refs.~\cite{PHLTA.B93.429,PHLTA.B303.334,PHLTA.B400.379,NUPHA.B710.485}. The collinear
counter terms at three loops are also known and have been obtained by the authors of Refs.~\cite{JHEPA.1211.062,PHLTA.B721.244,JHEPA.1310.096}. The three-loop virtual corrections are under control,
as they can be extracted from the three-loop quark and gluon form factors computed in
Refs.~\cite{0902.3519,1001.2887,1004.3653,1010.4478}. For finite top quark mass, an estimate of the \n3lo cross section for Higgs production was made in Ref.~\cite{1303.3590} by arguing that one can exploit
information encoded in the soft gluon and high energy limits to construct an accurate approximation to the full result.

In recent years, rapid progress has been made towards the analytical calculation of the relevant phase space integrals by expanding them
in the threshold limit\footnote{It should be noted that, as pointed out in Ref. \cite{1403.4616}, the evaluation of the leading term in the threshold expansion by itself is not completely satisfactory. It may be necessary in future work
to compute subleading terms in the threshold expansion as well to determine the size and impact of power corrections.}.
Two-loop, single-soft currents for Higgs and Drell-Yan production were computed to $\Ord (\e^0)$ in
Refs.~\cite{JHEPA.1202.056,hep-ph/0405236}, to $\Ord (\e^2)$ in
Ref.~\cite{1309.4391}, and, finally, to all orders in $\e$ in
Ref.~\cite{1309.4393}, where $\e$ is the usual parameter of dimensional regularization, $D = 4 - 2\e$.
The contribution from the square of the one-loop amplitude for Higgs boson production in association with a single soft parton is also known to all orders in $\e$~\cite{PHRVA.D60.116001,hep-ph/0007142,JHEPA.1312.088,1312.1296}.
Last year, the phase space integrals for soft triple-emission were
calculated in Ref.~\cite{1302.4379} to sufficiently high orders in $\e$ for an \n3lo calculation, and very recently, the phase space
integrals for Higgs boson production in association with two soft partons were computed by the same group, thereby completing the
threshold corrections to Higgs boson production through gluon fusion at
\n3lo~\cite{1403.4616}. However, this final piece of the soft part of the three-loop bare threshold cross section has not yet appeared in a separate publication. In this paper, we report on a parallel
calculation of the eikonal double-emission phase space integrals relevant to the \n3lo threshold calculation of gluon fusion Higgs or Drell-Yan lepton
production. Our results are remarkably compact and valid to all
orders in the dimensional regularization parameter. The higher order in $\e$ terms will be useful components of future N$^4$LO calculations, for example higher order  extractions of quark and gluon collinear anomalous dimensions.

We describe our calculational strategy below. At a hadron collider with incoming hadrons $N_1$ and $N_2$ and center-of-mass energy $\sqrt s$, the inclusive cross section for producing a Higgs boson of mass $M_\mathrm{H}$
or Drell-Yan lepton pair of invariant mass $M_\mathrm{DY}$ can be written as
\begin{eqnarray}
\label{eq:fulltheorycross}
  &&\sigma^{i} \left(s,M^2_i\right) =
  \nn \\
  &&\sum_{a,b} \int^1_0\! \dd x_1\, \dd x_2\,
  f_{a/N_1} \left(x_1, \mu^2_F\right)   f_{b/N_2} \left(x_2, \mu^2_F\right) \int^1_0 \! \dd
  z\, \delta \left( z - \frac{M^2_i}{\hat s}  \right)
  \hat{\sigma}_{ab}^{i} \left(\hat s, z, \alpha_s(\mu_R)\right)\,,
\end{eqnarray}
where $i=\mathrm{DY}$ or $\mathrm{H}$ and the summation is over all partonic channels. To save space, we will often use the abbreviations $\mathrm{DY}$ and $\mathrm{H}$ when discussing the processes that we consider in this paper. 
The parton distribution function of parton $n$ in hadron $N$, $f_{n/N}\left(x,\mu_F^2\right)$, depends on the momentum fraction, $x$, and the factorization
scale, $\mu_F$.  $\hat \sigma^{i}_{ab} (\hat{s}, z, \alpha_s(\mu_R))$, the partonic cross section, depends on the partonic center-of-mass energy, $\sqrt{\hat s}=\sqrt{x_1 x_2 s}$, the parameter $z$ defined by Eq. (\ref{eq:fulltheorycross}), and
the renormalized strong coupling constant at renormalization scale $\mu_R$, $\alpha_s(\mu_R)$. In this work, we take $\mu_F = \mu_R = M_i$ for the sake of simplicity.
In full fixed-order calculations, the dependence of partonic cross sections on $\hat{s}$ and $\als(M_i)$ is simple but their dependence on $z$ is quite non-trivial. Throughout this paper, we shall be concerned only with
a threshold expansion in the $z\to 1$ limit of the $\hat{\sigma}_{ab}^{i} \left(\hat s, z, \alpha_s(M_i)\right)$.
In this regime, additional partons radiated into the final state are constrained to be soft by virtue of the phase space constraint $\hat{s} \approx M^2_i$ and the partonic cross sections for both $\mathrm{H}$ and $\mathrm{DY}$
simplify dramatically:
\begin{eqnarray}
\label{eq:ggsimpcross}
  \hat{\sigma}^{\mathrm{H}}_{gg} (\hat s, z\rightarrow 1, \alpha_s(M_{\mathrm{H}})) &=& \sigma^{g g}_{0}(M_{\mathrm{H}}) \, G^{\mathrm{H}}(z)
\\
\label{eq:qqsimpcross}
  \hat{\sigma}^{\mathrm{DY}}_{q\bar{q}} (\hat s, z\rightarrow 1, \alpha_s(M_{\mathrm{DY}})) &=& \sigma^{q\bar{q}}_{0}(M_{\mathrm{DY}}) \, G^{\mathrm{DY}}(z) \,.
\end{eqnarray}
For simplicity, we keep the running coupling dependence of the functions introduced on the right-hand side of Eqs. (\ref{eq:ggsimpcross}) and (\ref{eq:qqsimpcross}) implicit. 
In $D = 4 - 2\e$ dimensions, $\sigma^{g g}_{0}(M_{\mathrm{H}})$ and $\sigma^{q\bar{q}}_{0}(M_{\mathrm{DY}})$ are given respectively
by\footnote{We follow Moch and Vogt \cite{hep-ph/0508265}  and write Eq.~(\ref{eq:Born}) with off-shell photons in mind.}
\begin{equation}
  \sigma^{g g}_{0}(M_{\mathrm{H}}) =  \frac{\pi \lambda^2(\als(M_{\mathrm{H}}))}{ 8 (1-\e) (N_c^2 - 1)} \qquad
  {\rm and} \qquad\sigma^{q\bar{q}}_{0}(M_{\mathrm{DY}}) = \frac{4\pi^2 (1-\e) \alpha(M_{\mathrm{DY}}) \,e_q^2}{N_cM^2_\mathrm{DY}} \, ,
\label{eq:Born}
\end{equation}
where $\lambda(\als(M_i))$ is the effective coupling of the Higgs boson to gluons in the limit of
infinite top quark mass~\cite{Shifman:1979eb},
\begin{equation}
  \mathcal{L}_{\mathrm{int}} = - \frac{1}{4}\lambda(\als(M_{\mathrm{H}})) H G^{\mu\nu,\,a}G^a_{\mu\nu}\, ,
\end{equation}
$N_c$ is the number of colors, $\alpha(M_{\mathrm{DY}})$ is the fine structure constant at renormalization scale $M_{\mathrm{DY}}$, and $e_q$ is the electric charge
of quark $q$. It is worth pointing out that $\lambda(\als(M_{\mathrm{H}}))$ has a perturbative expansion in the renormalized strong coupling constant of its own (see Ref. \cite{NUPHA.B744.121} for explicit expressions in
a slightly different normalization) and, throughout this work, we employ the conventional dimensional regularization scheme.
The coefficient functions $G^{\rm H}(z)$ and $G^{\rm DY}(z)$ contain singular distributions of the form
\begin{equation}
\label{eq:plusdistdef1}
  \delta(1-z) \qquad \mathrm{and} \qquad \left[\frac{\ln^k(1-z)}{1-z}
  \right]_+\ ,
\end{equation}
where the plus distribution $[g(z)]_+$ acts on functions regular in the $z \rightarrow 1$ limit as
\begin{equation}
\label{eq:plusdistdef2}
  \int^1_0 \! \dd z\, f(z) \left[g(z)\right]_+ = \int^1_0 \! \dd z\, \Big(f(z) -
  f(1)\Big) g(z)\,.
\end{equation}

The functional form of the coefficient
functions can be obtained by taking the $z\to 1$ limit of the full
partonic cross section. Alternatively, they can be directly calculated
in the soft limit of QCD, using a factorization formula derived from the
soft-collinear effective theory of QCD~\cite{hep-ph/0005275,hep-ph/0011336,hep-ph/0109045,hep-ph/0206152}. At our chosen scale, the $G^{i}(z)$ factorize into
products of the form~\cite{hep-ph/0605068,0710.0680,0808.3008}
\begin{equation}
\label{eq:factformula}
  G^{i}(z) = H^{i} \bar{S}^{i}(z)\,
\end{equation}
where the $H^{i}$ are hard functions encoding the virtual corrections at threshold and the $\bar{S}^{i}\left(z\right)$ are renormalized soft functions encoding the real radiative corrections at threshold.
At first, it might seem peculiar that we have not taken the soft functions in Eq. (\ref{eq:factformula}) to be a functions of $\omega/M_i$, where $\omega$ is twice the energy of the soft QCD radiation in the final state.
However, in threshold kinematics, we have
\begin{equation}
\label{eq:omegathreshold}
\omega = (1-z) \sqrt{\hat{s}} \approx (1-z) M_i
\end{equation}
and we see that the form of Eq. (\ref{eq:factformula}) follows naturally from Eq. (\ref{eq:omegathreshold}).

In general, hard functions in soft-collinear effective theory are complex squares of hard Wilson coefficients obtained by matching QCD onto soft-collinear effective theory.
The perturbative expansions of the Wilson coefficients relevant to Higgs and Drell-Yan production are known to the order required for an \n3lo calculation of the $G^{i}(z)$.
They can be extracted from the three-loop quark and gluon form factors~\cite{0902.3519,1001.2887,1004.3653,1010.4478}. In fact, they are written down explicitly in Ref. \cite{1004.3653}, Eqs. (7.4) - (7.9).
The actual hard coefficients that we need for our analysis are derived by taking the complex square of Eq. (7.3) in that
work, setting $\mu = M_i$, and then expanding in $\alpha_s(M_i)$ to third order. 

The bare soft functions can be written as squares of time-ordered matrix elements of pairs of semi-infinite Wilson line operators,
\begin{eqnarray}
\label{eq:softdef}
 S^{i}\left(z\right) = \frac{M_i}{d_i} \sum_{\mbox{\tiny $X_s$ }}
\langle 0 | T\Big\{Y_{n} Y_{\bar{n}}^\dagger\Big\} \delta\Big(\lambda - \hat{\bf P}^0\Big) | X_s\rangle\langle X_s | T\Big\{ Y_{\bar{n}} Y_{n}^\dagger \Big\} | 0 \rangle\,.
\end{eqnarray}
As mentioned above, the soft functions defined in Eq. (\ref{eq:softdef}) depend on the ratio of twice the energy of the soft QCD radiation to $M_i$ ({\it i.e.} on $1-z$),
as well as the bare strong coupling constant $\alpha_s$. $d_i$ is a conventional normalization constant,
\begin{equation}
d_\mathrm{H} = N_c^2 - 1  \quad\qquad {\rm and}\quad\qquad d_\mathrm{DY} = N_c\,,
\end{equation}
depending on whether the soft Wilson lines are in the adjoint or the fundamental representation of $\mathfrak{su}(N_c)$. The summation in Eq. (\ref{eq:softdef}) is over all possible soft parton final
states, $|X_s\rangle$. The operator $\hat{\bf P}^0$ acts on the final state $|X_s\rangle$ according to
\begin{equation}
 \hat{\bf P}^0|X_s\rangle = 2 E_{X_s} |X_s\rangle\,,
\end{equation}
where $E_{X_s}$ is the energy of the soft radiation in final state $|X_s\rangle$. The Wilson line operators, $Y_{n}$ and $Y_{\bar{n}}^\dagger$, are respectively defined as in-coming, path-ordered $\left(\mathbf{P}\right)$
 and anti-path-ordered $\left(\overline{\mathbf{P}}\right)$ exponentials \cite{hep-ph/0412110},
\begin{eqnarray}
  Y_{n}(x) &=& \mathbf{P} \exp \left( i g \int^0_{-\infty} \dd s\, n \cdot A(n s+x) \right)
\nn
\\
  Y_{\bar{n}}^\dagger(x) &=& \overline{\mathbf{P}} \exp \left( -i g \int_{-\infty}^0  \dd s\, \bar{n} \cdot A
    (\bar{n}s+x) \right).
\end{eqnarray}
In the above, $A_\mu = A_{\mu}^a T^a$, where the $T^a$ are either adjoint (for $\mathrm{H}$) or fundamental (for $\mathrm{DY}$) $\mathfrak{su}(N_c)$ matrices.
As usual, $n$ and $\bar{n}$ are light-like vectors whose space-like components are back-to-back and determine the
beam axis. For a generic four-vector, $k^\mu$, we have
\begin{equation}
n \cdot k = k^+ \qquad {\rm and} \qquad \bar{n} \cdot k = k^-
\end{equation}
given the usual definitions $k^+ = k^0 + k^3$ and $k^- = k^0 - k^3$.

In the perturbative regime, we can calculate the bare soft functions order-by-order in $\als$. They admit $\als$ expansions of the form
\begin{equation}
\label{eq:baresoft}
S^{i}(z) = \,\delta(1-z)  + \sum_{L = 1}^\infty
\left(\frac{\als S_\e}{4\pi}  \right)^L \left(\frac{\mu^2}{M^2_i} \right)^{\e L}
\frac{1}{(1-z)^{1+2\e L} } S^{i}_{L} (\e)\, .
\end{equation}
In Eq. (\ref{eq:baresoft}), $S_\e$ is the usual $\overline{\rm MS}$ factor,
\begin{equation}
 S_\e = \left(4 \pi e^{-\gamma_E}\right)^\e \,,
\end{equation}
where $\gamma_E=0.5772\dots$ is the Euler constant and $\mu$ is the 't Hooft scale introduced from continuing the space-time dimension to $D=4-2\e$ dimension.
After renormalization in the $\overline{\rm MS}$ scheme, it will be replaced by the renormalization scale, $\mu_R = M_i$.
The factor $1/(1-z)^{1+2\e L}$ can be expanded in $\e$ in terms of the plus distributions introduced above (Eqs. (\ref{eq:plusdistdef1}) and (\ref{eq:plusdistdef2})),
\begin{align}
  \frac{1}{(1-z)^{1+2\e L}} = -\frac{\delta(1-z)}{2 \e L} +
    \sum_{m=0}^\infty (-2 \e L)^m \left[ \frac{\ln^m (1-z)}{1-z}\right]_+ \ .
\label{eq:plusexp}
\end{align}

The expansion coefficients of the bare soft functions, $S^{i}_L(\e)$, are functions of $\e$ and $N_c$ only. In the case of $\mathrm{DY}$, the one- and two-loop results are well-known and
were calculated to all orders in $\e$ in Ref.~\cite{hep-ph/9808389}. 
The structure of the eikonal approximation is such that, through two-loop order, the color degrees of freedom completely
factorize from the part of the bare expansion coefficients which encodes the non-trivial soft dynamics; by simply replacing $C_F$
everywhere with $C_A$, the results reported in~\cite{hep-ph/9808389}
can be carried over in a straightforward manner to the case of $\mathrm{H}$ as well\footnote{Here, $C_A$ and $C_F$ are the usual quadratic Casimir invariants, expressed in terms of $N_c$ as
\begin{equation}
C_A = N_c \qquad {\rm and} \qquad C_F = \frac{N_c^2 - 1}{2 N_c}\,.
\end{equation}
}.
At \n3lo, many of the contributions can be converted from the $\mathrm{DY}$ language to the $\mathrm{H}$ language in a similar way\footnote{For example, both the  
single-emission, two-loop real-virtual corrections computed in Refs. \cite{1309.4391,1309.4393,PHRVA.D60.116001,hep-ph/0007142,JHEPA.1312.088,1312.1296} and
the double-emission, one-loop real-virtual corrections discussed in the present paper fall into this category.}
but, overall, the situation at three loops is significantly more complicated than at one and two loops. Although it is reasonable to expect {\it some} correspondence 
by virtue of the fact that the cut eikonal diagrams contributing to $\mathrm{H}$ and $\mathrm{DY}$ differ only in the Lie algebra representation
of their gluon-soft Wilson line interactions, there are sufficiently many color structures for both $\mathrm{H}$ and $\mathrm{DY}$ at three-loop order that finding such a correspondence requires non-trivial analysis.
In fact, with dedicated effort, it is possible to use the known results for the single-emission, two-loop real-virtual corrections \cite{1309.4391,1309.4393,PHRVA.D60.116001,hep-ph/0007142,JHEPA.1312.088,1312.1296},
the result for the double-emission, one-loop real-virtual corrections given in this article, and
the available constraints from the non-Abelian exponentiation theorem \cite{PHLTA.B133.90,NUPHA.B246.231} to determine the triple-emission real contributions to $\mathrm{DY}$
from the analogous result for $\mathrm{H}$ reported recently in Ref. \cite{1302.4379}. In other words, the full \n3lo result for $\mathrm{DY}$ can be deduced from results available in the literature and the main result of this article.
This line of analysis was quite useful and facilitated additional non-trivial cross-checks on our calculation which we summarize towards the end of this paper.
In order to keep the length of this article manageable, we defer a detailed discussion of this part of our analysis to a separate publication which focuses on the triple-emission real contributions.

As mentioned briefly in the introduction and in the last paragraph, there are several types of real radiative corrections in the threshold limit at three-loop order involving the emission of one or more soft partons. To obtain
complete results for the $S^{i}_3(\e)$, one must compute soft single-emission, two-loop~\cite{1309.4391,1309.4393} and one-loop squared real-virtual corrections~\cite{PHRVA.D60.116001,hep-ph/0007142,JHEPA.1312.088,1312.1296},
soft triple-emission real corrections~\cite{1302.4379}, and soft double-emission, one-loop real-virtual corrections, which are not yet publicly available\footnote{In a very recent paper~\cite{1403.4616},
these contributions were computed, thereby completing the calculation of the \n3lo Higgs boson production cross section at threshold.}.
In what follows, we compute the soft double-emission contributions to the $S^{i}_3(\e)$, referred to hereafter as $S^{i:\,{\rm 2R-V}}_3(\e)$, and fill this last remaining gap in the literature.
\begin{figure}[!htp]
\begin{center}
  \includegraphics[width=.6\textwidth,height=.6\textwidth,keepaspectratio]{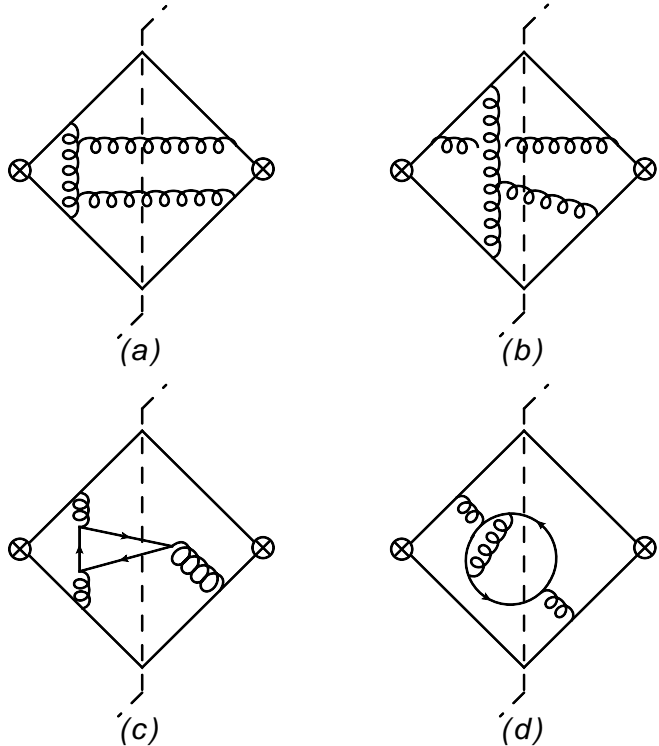}
  \caption{Panels $(a)$ and $(b)$ show cut eikonal Feynman diagrams with final state gluons and panels $(c)$ and $(d)$
  show cut eikonal Feynman diagrams with final state fermions.}
\label{fig:1}
\end{center}
\end{figure}

To generate the full set of cut eikonal Feynman diagrams, we use QGRAF~\cite{Nogueira:1991ex}. Representative cut diagrams are shown in Fig.~\ref{fig:1}. 
In order to test the correctness of our squared matrix elements while computing them, we found it convenient to use a general $R_\xi$ gauge for our internal gluons and lightcone
gauge for our final state gluons. As we explain below, the fact that our problem supplies two natural light-like reference vectors, $n$ and $\bar{n}$, allows for a very
stringent consistency check on our construction of the integrand. Once the diagrams are generated, they are processed by an in-house
{\tt FORM}~\cite{math-ph/0010025} code which dresses the cut eikonal diagrams with Feynman rules and
performs all relevant Dirac and color algebra. After that, we use an in-house {\tt Maple}
code to express all cut eikonal diagrams as a linear combination of Feynman integrals belonging to one of 28 integral families introduced for the eventual purpose of
performing an integration by parts reduction~\cite{PHLTA.B100.65,NUPHA.B192.159}
on the integrand using Laporta's algorithm~\cite{hep-ph/0102033}.
Mapping Feynman integrals to specific integral families is not entirely straightforward in this case and requires the application of numerous partial fraction identities to the raw integrand produced by our {\tt FORM} code.
This complication is due in part to the constraint on the kinematics imposed by the delta function
in the operator definition of the soft functions (see Eq. (\ref{eq:softdef})). We perform the integration by parts reduction using the development version of {\tt Reduze 2} \cite{1201.4330} which,
among other new features, supports the reduction of phase space integrals. The idea behind this was worked out some time ago and is usually called the reverse unitarity method~\cite{hep-ph/0207004,hep-ph/0306192,hep-ph/0312266}.
The key insight is that, for the purpose of integral reduction, one can use the relation
\begin{equation}
\delta\left(k^2\right) = -\frac{1}{2\pi i} \left( \frac{1}{k^2+i0} - \frac{1}{ k^2 - i0} \right)
\end{equation}
to replace delta function constraints with propagator denominators. After integration by parts reduction, the result is a rather simple linear combination of nine master integrals. 

To present our result in a compact fashion, let us first
introduce some additional notation. Besides the Casimir invariants $C_A$ and $C_F$, already defined above,
and the number of massless flavors, $n_f = 5$, our result for $S^{i:\,2{\rm R-V}}_{3}(\e)$ will depend on an additional Casimir invariant which itself depends on the index $i$:
\begin{equation}
C_\mathrm{H} = C_A \qquad {\rm and} \qquad C_\mathrm{DY} = C_F\,.
\end{equation}
Finally, if we make the definitions
\begin{equation*}
\int \left[k_1 k_2 q\right] = - i \pi^{3\e - 4} e^{3\gamma_E \e}\int \dd^D k_1\int \dd^D k_2\int \dd^D q\, \delta_+\left(k^2_1\right) \delta_+\left(k^2_2\right) \delta \Big( 1 - (n + \bar{n}) \cdot  (k_1 + k_2) \Big)
\end{equation*}
\begin{equation}
\begin{aligned}
\cd_1 &= q^2 + i 0^+\qquad
\\
\cd_3 &= (k_1 + k_2 - q)^2 + i 0^+\qquad
\\
\cd_5 &= 2 q \cdot n + i 0^+\qquad
\\
\cd_7 &= 2 k_1 \cdot \bar{n} + i 0^+\qquad
\\
\cd_9 &= 2 (k_1 - q) \cdot \bar{n}  + i 0^+\qquad
\end{aligned}
\begin{aligned}
\qquad\cd_2 &= (k_1 - q)^2 + i 0^+
\\
\qquad\cd_4 &= (k_1 + k_2 )^2 + i 0^+
\\
\qquad\cd_6 &= 2 k_1 \cdot n + i 0^+
\\
\qquad\cd_8 &= 2 k_2 \cdot \bar{n} + i 0^+
\\
\qquad\cd_{10} &= 2 (k_1 + k_2 - q) \cdot \bar{n}  + i 0^+\,,
\end{aligned}
\end{equation}
the nine master integrals mentioned above are given by
\begin{eqnarray}
\label{eq:firstmaster}
I_1(\e) &=& -12 (D-4) (D-3)^3 (3 D-11) {\rm Re}\left\{\int\left[k_1 k_2 q\right] \frac{1}{\cd_1 \cd_3}\right\}
\nn\\
&=&  \frac{e^{3 \gamma_E \e}\Gamma^4 ( 1 - \e)  \Gamma^2 (1 + \e) \Gamma ( 1 - 3 \e)}{ \Gamma^2 ( 1 - 2 \e) \Gamma(1+2\e)\Gamma ( 1 - 6 \e )} 
\nn\\
&=& 1-\frac{33 \zeta_2}{2}\epsilon ^2-65 \zeta _3 \epsilon^3+\frac{417 \zeta_4}{16} \epsilon^4 + \left(\frac{2145 \zeta _2 \zeta _3}{2}-\frac{7563\zeta_5}{5}\right) \epsilon ^5
\nn\\
&+& \left(-\frac{241541 \zeta _6}{128}+\frac{4225\zeta_3^2}{2}\right) \epsilon ^6 +\mathcal{O}\left(\e^7\right)
\\
I_2(\e) &=& 12 (D-4)^2 (D-3) (3 D-11) (3 D-10) {\rm Re}\left\{\int\left[k_1 k_2 q\right] \frac{1}{\cd_2 \cd_5 \cd_{10}} \right\}
\nn\\
&=& \frac{e^{3 \gamma_E \e} \Gamma^3 (1+ \e)\Gamma^2 (1-\e) \Gamma (1-2 \e)}{\Gamma(1+2\e)\Gamma(1-6 \e)}
\nn\\
&=& 1-\frac{31 \zeta _2 }{2}\epsilon ^2-67 \zeta _3 \epsilon ^3-\frac{359 \zeta_4 }{16}\epsilon ^4+\left(\frac{2077 \zeta _2 \zeta _3}{2}-\frac{7713\zeta _5}{5}\right) \epsilon ^5
\nn \\
&+&\left(-\frac{222195 \zeta _6}{128}+\frac{4489 \zeta_3^2}{2}\right) \epsilon ^6+\Ord\left(\epsilon^7\right)
\\
I_3(\e) &=& 2 (D-4)^4 (3 D-11)  {\rm Re}\left\{ \int [k_1 k_2 q] \frac{1}{ \cd_1 \cd_2 \cd_5 \cd_9 } \right\}
\nn\\
&=& \frac{e^{3 \gamma_E \e} \Gamma^4 (1-\e) \Gamma^3 (1+\e)}{\Gamma(1+2\e)\Gamma(1-6\e) }
\nn\\
&=& 1-\frac{33 \zeta _2 }{2}\epsilon ^2-69 \zeta _3 \epsilon ^3+\frac{225 \zeta_4 }{16}\epsilon ^4+\left(\frac{2277 \zeta _2 \zeta _3}{2}-\frac{7743 \zeta _5}{5}\right) \epsilon ^5
\nn \\
&+&\left(-\frac{209989 \zeta _6}{128}+\frac{4761 \zeta_3^2}{2}\right) \epsilon^6+\Ord\left(\epsilon^7\right)
\\
I_4(\e) &=& -2 (D - 4)^3 (D - 3) (3 D - 11){\rm Re}\left\{ \int [k_1 k_2 q] \frac{1}{\cd_1 \cd_3 \cd_6 \cd_8}\right\}
\nn\\
&=& \frac{e^{3 \gamma_E  \epsilon } \Gamma^4(1-\epsilon ) \Gamma^2 (1+\epsilon)\Gamma (1-3 \epsilon )   \,_3F_2(1,-\epsilon ,\epsilon ;1-2 \epsilon,1-\epsilon ;1)}{\Gamma^2 (1-2 \epsilon )\Gamma(1+2\e)\Gamma (1-6 \epsilon ) }
\nn\\
&=& 1-\frac{35 \zeta _2 }{2}\epsilon ^2-71 \zeta _3 \epsilon ^3+\frac{713 \zeta_4 }{16}\epsilon ^4+\left(\frac{2473 \zeta _2 \zeta _3}{2}-\frac{7938\zeta _5}{5}\right) \epsilon ^5
\nn\\
&+&\left(-\frac{193303 \zeta _6}{128}+\frac{5005 \zeta_3^2}{2}\right) \epsilon^6+\Ord\left(\epsilon^7\right)
\\
I_5(\e) &=& (D-4)^3 (D-3) (3 D-11){\rm Re}\left\{ \int[k_1 k_2 q] \frac{1}{ \cd_1 \cd_3 \cd_5 \cd_{10}} \right\}
\nn\\
&=& \frac{e^{3 \gamma_E  \epsilon } \epsilon ^2 
   \Gamma^5 (1-\epsilon ) \Gamma^3 (1+\epsilon) \Gamma (1-3 \epsilon ) \, _3F_2(1,1+\epsilon,1+2 \epsilon;2-\epsilon ,2+\epsilon;1)}{\Gamma^2 (1-2 \epsilon )\Gamma(1+2\e)\Gamma (1-6 \epsilon )  \Gamma(2-\epsilon ) \Gamma (2+\epsilon)}
\\
&=& \zeta _2 \epsilon ^2+3 \zeta _3 \epsilon ^3-29 \zeta _4 \epsilon
   ^4+\left(-\frac{229 \zeta _2 \zeta _3}{2}+\frac{75 \zeta _5}{2}\right)
   \epsilon ^5+\left(-\frac{12155 \zeta _6}{64}-195 \zeta _3^2\right) \epsilon
   ^6+\Ord\left(\epsilon ^7\right)
\nn\\
I_6(\e) &=& 2 (D-4)^3 (2 D-7) (3 D-11) {\rm Re}\left\{\int [k_1 k_2 q] \frac{1}{\cd_2 \cd_4 \cd _5 \cd_{10}} \right\}
\nn\\
&=& \frac{e^{3 \gamma_E \e} \Gamma^3 (1+ \e)
  \Gamma^2 (1 - \e) \Gamma^2 (1-3 \e)\, _3F_2( -\e,\e,\e; 1 - 2 \e, 1 - \e; 1 )  }{\Gamma(1+2\e)\Gamma (1 - 4 \e)\Gamma (1-6 \e)} 
\nn\\
&=& 1-\frac{33 \zeta _2 }{2}\epsilon ^2-74 \zeta _3 \epsilon ^3-\frac{223 \zeta
   _4 }{16}\epsilon ^4+\left(1222 \zeta _2 \zeta _3-\frac{16701 \zeta
   _5}{10}\right) \epsilon ^5
\nn\\
&+&\left(-\frac{167333\zeta _6}{128}+\frac{5483 \zeta _3^2}{2}\right) \epsilon ^6+\Ord\left(\epsilon ^7\right)
\\
I_7(\e) &=& \frac{3 (D-4)^5}{10} {\rm Re}\left\{ \int [k_1 k_2 q] \frac{1}{ \cd_1 \cd_2 \cd_3 \cd_6 \cd_{10} } \right\}
\nn\\
&=&   \frac{2 e^{3 \gamma_E  \e} \Gamma^4 (1-\e)\Gamma^2(1+\e) \Gamma (1-3 \e)\, _3F_2 (1,1,-3 \e; 1-3 \e,1-2 \e; 1)}{5 \Gamma^2(1-2\e)\Gamma(1+2\e)\Gamma(1-6\e) }
\nn\\
&=& 1-\frac{201 \zeta _2 }{10}\epsilon ^2-\frac{451 \zeta _3 }{5}\epsilon^3+\frac{5757 \zeta _4 }{80}\epsilon ^4+\left(\frac{17439 \zeta _2
   \zeta _3}{10}-\frac{9489 \zeta _5}{5}\right) \epsilon ^5
\nn\\
&+&\left(-\frac{163813 \zeta _6}{640}+\frac{38261
   \zeta _3^2}{10}\right) \epsilon^6+\Ord\left(\epsilon ^7\right)
\\
I_8(\e) &=& \frac{3 (D-4)^5 (3 D-14)}{11 (3 D-13)} {\rm Re}\left\{\int [k_1 k_2 q] \frac{1}{ \cd_2 \cd_3 \cd_4 \cd_5 \cd_6 \cd_{10} } \right\}
\nn\\
&=&  \frac{9 e^{3 \gamma_E \e} \Gamma^3 (1 + \e) \Gamma^2 (1 - \e)  \Gamma ( 1 - 2 \e)\, _4F_3( 1, -2 \e, - \e, 1 + \e; 1 - \e, 1 - \e, -3 \e; 1 )}{11 \Gamma(1+2\e)\Gamma (1 - 6 \e)} 
\nn\\
&=& 1-\frac{349 \zeta _2 }{22}\epsilon ^2-\frac{741 \zeta _3 }{11}\epsilon^3-\frac{1197 \zeta _4 }{176}\epsilon ^4+\left(\frac{23427 \zeta _2 \zeta _3}{22}-\frac{84093 \zeta _5}{55}\right) \epsilon^5
\nn\\
&+&\left(-\frac{2443225 \zeta _6}{1408}+\frac{49683 \zeta _3^2}{22}\right)  \epsilon ^6+\Ord\left(\epsilon ^7\right)
\\
\label{eq:lastmaster}
I_9(\e) &=& \frac{9 (D-4)^5 (3 D-14)}{16 (3 D-13)} {\rm Re}\left\{\int [k_1 k_2 q] \frac{1}{ \cd_1 \cd_3 \cd_4 \cd_ 5 \cd_6 \cd_8 \cd_{10} } \right\}
\nn\\
&+&\frac{1}{3 D - 14} \Big(-2 I_1(\e) + 3 I_4(\e) - 6 I_5(\e)\Big)
\nn\\
&=& \frac{e^{3 \gamma_E  \epsilon }  \Gamma^3(1-\epsilon ) \Gamma^3 (1+\epsilon)\Gamma^2 (1-3 \epsilon )}{2  \Gamma (-6 \epsilon)\Gamma(1+2\e)\Gamma(1-2\e) \Gamma (1-4 \epsilon )\Gamma (2+\epsilon)}\times
\nn \\
&\times&F^{0:3:3}_{1:1:1}\left(\begin{array}{ccc} \mbox{{\bf ---}}~ &1,1,-2\e~ &-\e,-\e,-2\e \\ -4\e~ &2+\e~ &1-\e\end{array}; 1, 1\right)
\nn\\
&-&\frac{1}{2(1 + 3 \e)} \Big(-2 I_1(\e) + 3 I_4(\e) - 6 I_5(\e)\Big)
\nn\\
&=& 1 - 3 \zeta_2 \epsilon^2 - 14 \zeta_3 \epsilon ^3 - \frac{5235 \zeta_4}{16} \epsilon^4+  \left(-\frac{1455 \zeta_2 \zeta_3}{2}-\frac{7131 \zeta_5}{10}\right)\epsilon^5
\nn\\
&+&  \left(-\frac{312085 \zeta_6}{64}-\frac{2747 \zeta_3^2}{2}\right)\epsilon ^6+\Ord\left(\epsilon ^7\right)\,.
\end{eqnarray}
While integrals $I_1(\e)$ through $I_8(\e)$ can be simply expressed in terms of gamma functions and generalized hypergeometric functions, $I_9(\e)$ is more complicated. 
In particular, it involves the Kamp\'{e} de F\'{e}riet function (see Ref. \cite{Exton}, Eq. (1.3.2.1) for our Kamp\'{e} de F\'{e}riet function conventions)
\begin{eqnarray}
F^{0:3:3}_{1:1:1}\left(\begin{array}{ccc} \mbox{{\bf ---}}~ &1,1,-2\e~ &-\e,-\e,-2\e \\ -4\e~ &2+\e~ &1-\e\end{array}; 1, 1\right) &=&\sum_{m = 0}^\infty \sum_{n = 0}^\infty
\frac{m! (-2\e)_m \left[(-\e)_n\right]^2(-2\e)_n}{n! (1-\e)_n (2+\e)_m (-4\e)_{n+m}}\,,
\nn\\
{\rm where}~~(r)_s &=& \frac{\Gamma(r+s)}{\Gamma(r)}\,. 
\end{eqnarray}

A few words about our derivation of the master integrals are in order. Although, for the most part, we wish to defer the discussion of the novel method used to do the integrals
to a more technical future publication, let us briefly describe some consistency checks on Eqs.~(\ref{eq:firstmaster})-(\ref{eq:lastmaster}) that we found useful. For all integrals, a good first step is to integrate out the virtual momentum $q$
since, in all cases, one obtains simple expressions built out of gamma functions and Gauss hypergeometric functions.
The numerical sector decomposition~\cite{hep-ph/0004013,0709.4092} code {\tt FIESTA 3}~\cite{1312.3186} was successfully used to check our virtual integrations for sign errors.\footnote{It is worth pointing out that, although we have
absorbed the operation of taking the real part into the definitions of our master integrals, they are complex functions of $\e$ at the level of {\tt FIESTA} and it is therefore essential to use version 3 since earlier versions of the code support
only Euclidean kinematics.} Unfortunately, more stringent checks on the virtual integrations with {\tt FIESTA 3} could not be performed with confidence due to apparent stability issues with the numerical integration routines.
For all of the integrals except $I_5(\e)$ and $I_9(\e)$, we found that the phase space integrals over $k_1$ and $k_2$ that remain once $q$ has been integrated out could be numerically checked in {\tt Mathematica}
for appropriate real, negative values of $\e$ using a parametrization employed in earlier work (see Refs. \cite{1105.3676,1112.3343,1309.3560}). In all cases where our parametrization applies\footnote{Roughly speaking, we expect 
our methods to apply in a straightforward manner if the dependence of the phase space integrals on the dot product 
$k_1 \cdot k_2$ is of the form $(k_1\cdot k_2)^{p(\e)}$ for some linear polynomial, $p(\e)$.}, we found that our analytical results could be checked
numerically at the level of the phase space integrals to at least seven significant digits for several appropriately chosen
but essentially random values of $\e$.\footnote{Of course, values of $\e$ which lead to a convergent numerical integration must be chosen.}
Fortunately, the two integrals whose phase space integrals do not admit a straightforward parametrization for technical reasons, $I_5(\e)$ and $I_9(\e)$,
are actually the simplest to set up in our non-standard analytical approach. Our results for $I_5(\e)$ and $I_9(\e)$ were verified
{\it a posteriori} by using them to carry out the stringent global consistency checks described at the end of this work. To derive the Taylor expansions of integrals $I_1(\e)$ through $I_8(\e)$,
we made extensive use of the {\tt Mathematica} package {\tt HypExp} \cite{hep-ph/0507094,0708.2443}.  As a sanity check, we performed a completely independent expansion of
the master integral $I_6(\e)$ using general-purpose integration scripts written by one of us (see Ref. \cite{1309.3560}) and found complete agreement.

The master integral $I_9(\e)$ is somewhat more subtle than $I_1(\e)$-$I_8(\e)$. In an attempt to expand the second line of Eq. (\ref{eq:lastmaster}) in $\e$, we encountered the single-parameter integral
\begin{align}
  \int^{1}_0 \! \dd u \,u^{-2-3\e}(1-u)^{-1-2\e} \Big({}_2F_1\left(-\e,-\e; 1-\e; 1-u\right)\Big)^2\,.
\label{eq:illdef}
\end{align}
Historically, such integrals were considered pathological due to the fact that, na\"{i}vely, they do not appear to be sector-decomposable. The problem is that the integrand of (\ref{eq:illdef}) has small $u$ asymptotics of the 
too-singular form $u^{-2-3\e}$. Integrals such as (\ref{eq:illdef}) have appeared in other QCD computations. For example, in Ref. \cite{hep-ph/0404293}, a dedicated unitarity-based sewing method
was used to avoid classes of integrals with similar power-law singularities. In fact, we found that we were able to treat integrals like (\ref{eq:illdef}) in a very direct way.
First, observe that the above integral converges for all complex $\e$ such that $\mathrm{Re}(\e) < -1/3$.
In spite of the fact that the integral does not converge in a neighborhood of $\e = 0$, one can expand (\ref{eq:illdef}) in a Laurent series about $\e = 0$ anyway by carefully applying the principle of analytical continuation.
To understand this, begin by considering a fictitious, alternative representation of our result, Eq. (\ref{eq:integrand}), written in terms of integrals free of power-law divergences which converge in a neighborhood of $\e = 0$.
This hypothetical representation should always exist if all of the divergences in the calculation are regulated by $\e$. Using the fact that all Feynman integrals in dimensional regularization are analytic functions of $\e$,
one can perform an analytical continuation to a value of $\e$ which happens to simultaneously lie in the region of convergence of (\ref{eq:illdef}) and $I_1(\e)$ through $I_8(\e)$. Now, by performing an integration by parts reduction on the fictitious representation, 
one can recover the precise form of Eq. (\ref{eq:integrand}), written in terms of our preferred basis of master integrals. Due to the fact that, by assumption,
the point in the complex $\e$ plane to which the original analytical continuation was performed lies within the 
domain of convergence of all of our master integrals, one can derive the closed formulas for $I_1(\e)$-$I_9(\e)$ given above (the second lines of Eqs. (\ref{eq:firstmaster})-(\ref{eq:lastmaster})) in this region of the complex $\e$ plane.
Finally, one can analytically continue back to the point $\e = 0$ and derive Laurent expansions for all of the masters integrals
using standard techniques. Carrying out this program explicitly was necessary to Laurent expand $I_9(\e)$ about $\e = 0$ and our analysis will be presented in a future publication which features an in-depth discussion of the master integrals.

Our choice of integral basis also requires some explanation.
As is clear from the Taylor series expansions of our master integrals, the definitions we
have made are such that all of the $I_j(\e)$ are pure functions in the sense of reference \cite{1304.1806}. As we shall see, the
all-orders-in-$\e$ result for $S^{i:\,2{\rm R-V}}_{3}(\e)$ assumes a particularly simple form when expressed in terms of the integral basis defined above.
After integration by parts reduction, it can be written as
\begin{eqnarray}
\label{eq:integrand}
S^{i:\,2{\rm R-V}}_{3}(\e) &=& \crr \ca^2 \left[
\left(-\frac{170512}{243 (D-4)}+\frac{6368}{9(D-3)}-\frac{560}{243 (D-1)}+\frac{14272}{27(D-4)^2}+\frac{180}{(D-3)^2}
\breakll
-\frac{68}{81 (D-1)^2}-\frac{8576}{27(D-4)^3}+\frac{1408}{9 (D-4)^4}+\frac{256}{27 (D-4)^5}-\frac{32}{9(D-6)}\right) I_1(\e)
\breakl
+\left(-\frac{160000}{243 (D-4)}+\frac{2048}{3(D-3)}-\frac{32}{D-2}+\frac{12032}{1215 (D-1)}+\frac{52864}{81(D-4)^2}
\breakll
-\frac{17152}{27 (D-4)^3}+\frac{5632}{9 (D-4)^4}-\frac{2816}{9(D-4)^5}-\frac{32}{15 (D-6)}\right) I_2(\e)+\frac{768}{(D-4)^5}I_3(\e)
\breakl
+\left(\frac{160}{9(D-4)}-\frac{64}{3 (D-3)}+\frac{64}{45 (D-1)}-\frac{64}{3(D-4)^2}-\frac{256}{3(D-4)^5}
\breakll
+\frac{32}{15 (D-6)}\right) I_4(\e)+\left(-\frac{320}{9 (D-4)}+\frac{128}{3(D-3)}-\frac{128}{45 (D-1)}+\frac{128}{3(D-4)^2}
\breakll
+\frac{512}{3(D-4)^5}-\frac{64}{15(D-6)}\right) I_5(\e)+\left(\frac{81728}{243(D-4)}-\frac{1024}{3 (D-3)}+\frac{896}{1215 (D-1)}
\breakll
-\frac{28160}{81(D-4)^2}+\frac{8576}{27 (D-4)^3}-\frac{2816}{9 (D-4)^4}-\frac{512}{9(D-4)^5}+\frac{64}{15 (D-6)}\right) I_6(\e)
\breakl
- \frac{10240}{27 (D-4)^5}I_7(\e) - \frac{2816}{9 (D-4)^5}I_8(\e) -\frac{512}{9(D-4)^5}I_9(\e)
\right]
\nn
\\
&&
- \crr^2 \ca \frac{1536}{(D - 4)^5}I_3(\e)
\nn
\\
&&
+ \crr \ca \nf \left[ 
\left(-\frac{928}{243 (D-4)}-\frac{64}{9 (D-3)}+\frac{16}{D-2}-\frac{1664}{243(D-1)}-\frac{16}{9(D-1)^2}
\breakll
+\frac{1504}{81 (D-4)^2}-\frac{16}{(D-3)^2}-\frac{1216}{27 (D-4)^3}+\frac{256}{9 (D-4)^4}+\frac{16}{9(D-6)}\right)I_1(\e)
\breakl
+\left(\frac{27328}{243 (D-4)}-\frac{512}{3 (D-3)}+\frac{112}{3(D-2)}+\frac{24064}{1215 (D-1)}-\frac{8320}{81(D-4)^2}
\breakll
+\frac{64}{(D-2)^2}+\frac{2560}{27 (D-4)^3}-\frac{1024}{9(D-4)^4}+\frac{16}{15 (D-6)}\right) I_2(\e)+\left(-\frac{256}{9(D-4)}
\breakll
+\frac{128}{3 (D-3)}-\frac{16}{D-2}+\frac{128}{45 (D-1)}+\frac{64}{3(D-4)^2}-\frac{16}{15 (D-6)}\right) I_4(\e)
\breakl
+\left(\frac{512}{9 (D-4)}-\frac{256}{3(D-3)}+\frac{32}{D-2}-\frac{256}{45 (D-1)}-\frac{128}{3(D-4)^2}
\breakll
+\frac{32}{15 (D-6)}\right) I_5(\e)+\left(-\frac{17984}{243(D-4)}+\frac{256}{3 (D-3)}-\frac{32}{3 (D-2)}-\frac{32}{15 (D-6)}
\breakll
+\frac{5888}{81 (D-4)^2}-\frac{1280}{27 (D-4)^3}+\frac{512}{9(D-4)^4}+\frac{1792}{1215(D-1)}\right) I_6(\e)\right]
\nn
\\
&&
+ \crr \cf \nf 
\left(-\frac{544}{3 (D-3)}+\frac{160}{243 (D-1)}-\frac{128}{3(D-3)^2}
\breakl
+\frac{43904}{243 (D-4)}-\frac{11072}{81 (D-4)^2}+\frac{2432}{27(D-4)^3}-\frac{512}{9 (D-4)^4}\right) I_1(\e)
\nn
\\
&&
+ \crr \nf^2 \left(-\frac{64}{3 (D-3)}-\frac{16}{3 (D-3)^2}-\frac{16}{81(D-1)^2}
\breakl
+\frac{64}{3 (D-4)}-\frac{1280}{81 (D-4)^2}+\frac{256}{27 (D-4)^3}\right) I_1(\e)\,.
\end{eqnarray}
The compact form of Eq. (\ref{eq:integrand}) is due in large part to the absence of spurious poles in the coefficients of our master integrals;
the coefficients of the basis integrals in Eq. (\ref{eq:integrand}) have poles at positive integer values of $D$ only. By rewriting the above expression in terms of a more conventional basis such as the one supplied by {\tt Reduze 2}
out of the box, it becomes clear that this intriguing feature of our result is a direct consequence of the fact that the $I_j(\e)$ are pure functions. For a problem with similar kinematics,
the computation of the three-loop form factors in massless QCD, it was observed in Ref.~\cite{1010.1334} (in slightly different language) that an integral basis of pure functions could be written down for all master integrals.
However, to the best of our knowledge, our paper is the first to show that, even for single-scale problems,
it is reasonable to expect the pole structure of the reduced integrand to dramatically simplify when it is expressed in terms of a basis of pure functions.
Substituting the Taylor expansions of the master integrals, Eqs. (\ref{eq:firstmaster})-(\ref{eq:lastmaster}), into Eq.~(\ref{eq:integrand}) and expanding in $\e$, we find
\begin{eqnarray}
S^{i:\,2{\rm R-V}}_{3}(\e) &=&  \crr \ca^2 \left[\frac{40}{3\e^5} + \frac{88}{3 \e^4} + \left(\frac{2144}{27}-\frac{692 \zeta _2}{3} \right) \frac{1}{\e^3} + \left(\frac{16448}{81}-\frac{4004 \zeta _2}{9}-\frac{3208 \zeta_3}{3}\right) \frac{1}{\e^2} 
\breakl
+ \left(\frac{40744}{81}-\frac{33232 \zeta_2}{27}-\frac{16280\zeta_3}{9}-\frac{667 \zeta_4}{6} \right) \frac{1}{\e} + \frac{870688}{729} -\frac{256880 \zeta_2}{81}
\breakl
-\frac{134000 \zeta_3}{27}-\frac{2101 \zeta_4}{6}+ 18380 \zeta_2 \zeta_3 -\frac{70744 \zeta_5}{3} + \left(\frac{6039424}{2187}-\frac{1930348 \zeta_2}{243}
\breakll
-\frac{114112\zeta_3}{9}-\frac{1742 \zeta_4}{9}+\frac{81620\zeta _2 \zeta _3}{3}-\frac{636944 \zeta_5}{15}-\frac{1591043 \zeta_6}{144}+\frac{125284 \zeta _3^2}{3} \right) \e\right]
\nn
\\
&&
+ \crr^2 \ca \left[\frac{48}{\e^5}  -  \frac{792 \zeta _2}{\e^3} -  \frac{3312 \zeta _3}{\e^2} +  \frac{675 \zeta _4}{\e} + 54648 \zeta _2 \zeta _3 
\breakl
-\frac{371664 \zeta_5}{5}+ \left(-\frac{629967 \zeta_6}{8} + 114264 \zeta _3^2 \right) \e \right]
\nn
\\
&&
+\crr \ca \nf \left[-\frac{16}{9\e^4}  -\frac{8}{27\e^3} + \left( \frac{200}{81}+\frac{200 \zeta_2}{9} \right)\frac{1}{\e^2} + \left( -\frac{752}{243}-\frac{188 \zeta_2}{27}
\breakll
 +\frac{880 \zeta_3}{9} \right) \frac{1}{\e}-\frac{22384}{729}-\frac{6676 \zeta _2}{81}-\frac{280 \zeta_3}{27}+\frac{469 \zeta_4}{3}+ \left(-\frac{251120}{2187}
\breakll
-\frac{11624 \zeta_2}{243}-\frac{27272 \zeta_3}{81}+\frac{5941 \zeta_4}{18}-\frac{3400}{3} \zeta _2 \zeta_3+\frac{11712 \zeta_5}{5} \right) \e\right]
\nn
\\
&&
+\crr \cf \nf \left[-\frac{32}{9\e^4} -\frac{304}{27\e^3} + \left( -\frac{2768}{81}+\frac{176 \zeta_2}{3} \right) \frac{1}{\e^2} + \left( -\frac{21952}{243}+\frac{1672 \zeta _2}{9}
\breakll
+\frac{2080 \zeta_3}{9} \right) \frac{1}{\e} -\frac{163136}{729} + \frac{15224 \zeta_2}{27}+\frac{19760\zeta_3}{27}-\frac{278 \zeta_4}{3}
\breakl
+ \left( -\frac{1166080}{2187}+\frac{120736 \zeta_2}{81}+\frac{179920 \zeta_3}{81}-\frac{2641 \zeta_4}{9}-\frac{11440}{3} \zeta _2 \zeta_3+\frac{80672 \zeta_5}{15} \right) \e\right]
\nn
\\
&&
+ \crr \nf^2 \left[-\frac{32}{27\e^3} -\frac{320}{81\e^2} + \left( -\frac{32}{3}+\frac{176 \zeta_2}{9} \right) \frac{1}{\e} - \frac{19456}{729} + \frac{1760 \zeta _2}{27}+\frac{2080 \zeta_3}{27} 
\breakl
 + \left(-\frac{140032}{2187} + 176 \zeta _2+\frac{20800 \zeta_3}{81}-\frac{278 \zeta_4}{9}\right) \e \right] + \mathcal{O}\left(\e^2\right)\ .
\label{eq:result}
\end{eqnarray}
The Laurent series expansion of $S^{i:\,2{\rm R-V}}_{3}(\e)$ given above is the main result of this paper.\\
$~~~~~$In order to ensure the correctness of our results, we found it useful to perform several
consistency checks at various stages of the calculation. First, recall that, while constructing the raw integrand, we made use of a physical lightcone gauge for our final state gluons. It turns out that running our code twice,
once with $n$ as the lightcone reference vector for all gluons and once with $\bar{n}$ as the lightcone reference vector for all gluons, allows for a very stringent check at the stage of integral reduction. The reason for this is 
as follows. At the level of the unreduced integrand, using a physical gauge with some choice of reference vector breaks the $n-\bar{n}$ symmetry of the expression expected by virtue of the fact that our soft Wilson line
operators are back-to-back. This symmetry, however, is restored after integral reduction and, most importantly, before reducing all Feynman integrals to masters, we observed that many integrals enter
one or the other of the two expressions we derived but not both. Thus, reducing our two seemingly different expressions for the raw integrand down to Eq. (\ref{eq:integrand})
furnished a non-trivial check on the calculation. This check was complementary to a more conventional check on the gauge invariance of Eq. (\ref{eq:integrand}); we employed a general $R_\xi$ gauge for our internal gluons
and confirmed that the gauge parameter, $\xi$, drops out of our final results.\\
$~~~~~$The final and most important check on our calculation was an explicit comparison to the relevant literature. After extracting the one-, two-, and three-loop
expansion coefficients of the hard functions from Refs.~\cite{0902.3519,1001.2887,1004.3653,1010.4478,NUPHA.B319.570,hep-ph/0007289} and deducing the one-, two-, and three-loop expansion coefficients
of the renormalized soft functions from our calculation and various complete and partial results existing in the literature \cite{hep-ph/9808389,1309.4391,1309.4393,PHRVA.D60.116001,hep-ph/0007142,JHEPA.1312.088,1312.1296,1302.4379},
we were able to use Eq. (\ref{eq:factformula}) to reproduce the full tower of plus distributions of the form
\begin{equation*}
\left[\frac{\ln^m(1-z)}{(1-z)}\right]_+, \quad\qquad m = 0\dots5
\end{equation*}
for both gluon fusion Higgs production and Drell-Yan lepton production. In this endeavor, we used~\cite{hep-ph/0508265} as our primary reference for the predictions provided by renormalization group invariance for the processes
of interest but should point out that 
similar studies have been carried out by many different groups over the years~\cite{hep-ph/0508284,hep-ph/0603041,hep-ph/0605068,0809.4283}.
Even better, for the case of Higgs boson production, we were able to reproduce the $\delta(1-z)$ terms in the full \n3lo result obtained recently by the authors of Ref.~\cite{1403.4616} and,
for the case of Drell-Yan lepton production, we were able to reproduce the
prediction for the $\delta(1-z)$ terms made recently in Ref.~\cite{1404.0366} using completely different methods.
Of course, this is really a check on the $\delta(1-z)$ term coming from the double-emission, one-loop real-virtual corrections treated in this paper, not the $\delta(1-z)$ term in the full \n3lo result because, so far,
no independent calculation of the triple-emission real contributions has appeared confirming that the analysis of \cite{1302.4379} is correct. Nevertheless, the fact
that we were able to completely reproduce the results presented in Ref.~\cite{1403.4616} is highly non-trivial and very encouraging.\\
$~~~~~$We have taken a decisive step in this work towards a fully independent calculation of the total cross section at threshold for gluon fusion Higgs production and Drell-Yan lepton pair production at \n3lo.
In this paper, we focused on the calculation of the double-emission, one-loop real-virtual contributions to the bare three-loop threshold soft functions for gluon fusion Higgs production and Drell-Yan lepton production.
As explained above, we found results completely consistent with all of the available literature on the subject. During the course of our calculation, we made significant technical advances in the evaluation of cut eikonal phase space integrals
which allowed us to compute all of our master integrals to all orders in $\e$. It would be very interesting to try and carry out a similar program for the master integrals that one finds in the computation of the triple-emission real contributions
to the bare three-loop soft functions. This would be useful because complete all-orders-in-$\e$ results for the three-loop bare soft functions would save future researchers the trouble
of revisiting the calculations performed in Ref. \cite{1302.4379} if it turns out
to be necessary to calculate to one order higher (to N$^4$LO) in perturbative QCD. In fact, it may well be that our master integrals find useful application as input integrals to the systems of differential equations that will presumably
be derived in the future when an attempt is made to calculate the full $z$-dependent partonic cross sections for the processes we have considered at threshold.\\
$~~~~~$We also noted in this work that the use of an integral basis composed of pure functions leads to remarkable simplifications at the level of the reduced integrand, Eq. (\ref{eq:integrand}).
In stark contrast to more conventional bases of master integrals, our basis of pure functions is such that the coefficients of the master integrals have poles only at $D = 1$, $2$, $3$, $4$, and $6$. 
We argued that this property is a direct consequence of the niceness
of our integral basis which would certainly not hold in general. In fact, the most complicated pure function in our basis was identified by demanding that spurious singularities at $D = 14/3$ disappear. In the near future, we plan to complete the
research program initiated by two of us in Ref. \cite{1309.4391} and treat the triple-emission real corrections.
Whether or not we will manage to obtain all-orders-in-$\e$ expressions remains to be seen and should help us to more accurately assess the scope of applicability
of the methods that we developed to carry out the calculations discussed in this paper. Our hope is that the \n3lo threshold production cross section calculations considered in this work and
elsewhere \cite{1309.4391,1309.4393,PHRVA.D60.116001,hep-ph/0007142,JHEPA.1312.088,1312.1296,1302.4379,1403.4616} will, once \n3lo parton distribution function fits become available, allow for
more accurate inclusive predictions than ever before and significantly advance the physics program of the Large Hadron Collider.

\begin{acknowledgments}

We are grateful to Sven-Olaf Moch for useful correspondence. The Feynman diagrams are drawn with JaxoDraw~\cite{hep-ph/0309015}, based on AxoDraw~\cite{CPHCB.83.45}.
The research of YL and HXZ is
supported by the US Department of Energy under contract
DE–AC02–76SF00515.
The research of AvM is supported in part by the
Research Center {\em Elementary Forces and Mathematical Foundations (EMG)} of the
Johannes Gutenberg University of Mainz and by the
German Research Foundation (DFG). The research of RMS is supported by the ERC
Advanced Grant EFT4LHC of the European Research Council, the Cluster of Excellence Precision Physics, Fundamental Interactions and Structure of Matter (PRISMA-EXC 1098).

\end{acknowledgments}


\begin{thebibliography}{99}
%\cite{NUPHA.B157.461}
\bibitem{NUPHA.B157.461} 
  G.~Altarelli, R.~K.~Ellis and G.~Martinelli,
  ``Large Perturbative Corrections to the Drell-Yan Process in QCD,''
  Nucl.\ Phys.\ B {\bf 157}, 461 (1979).
  %%CITATION = NUPHA,B157,461;%%
  %778 citations counted in INSPIRE as of 04 Apr 2014


%\cite{NUPHA.B359.283}
\bibitem{NUPHA.B359.283} 
  S.~Dawson,
  ``Radiative corrections to Higgs boson production,''
  Nucl.\ Phys.\ B {\bf 359}, 283 (1991).
  %%CITATION = NUPHA,B359,283;%%
  %630 citations counted in INSPIRE as of 04 Apr 2014


%\cite{NUPHA.B359.343}
\bibitem{NUPHA.B359.343} 
  R.~Hamberg, W.~L.~van Neerven and T.~Matsuura,
  ``A Complete calculation of the order $\alpha_s^2$ correction to the Drell-Yan $K$ factor,''
  Nucl.\ Phys.\ B {\bf 359}, 343 (1991)
  [Erratum-ibid.\ B {\bf 644}, 403 (2002)].
  %%CITATION = NUPHA,B359,343;%%
  %804 citations counted in INSPIRE as of 04 Apr 2014


%\cite{PRLTA.70.1372}
\bibitem{PRLTA.70.1372} 
  D.~Graudenz, M.~Spira and P.~M.~Zerwas,
  ``QCD corrections to Higgs boson production at proton proton colliders,''
  Phys.\ Rev.\ Lett.\  {\bf 70}, 1372 (1993).
  %%CITATION = PRLTA,70,1372;%%
  %266 citations counted in INSPIRE as of 04 Apr 2014


%\cite{hep-ph/9504378}
\bibitem{hep-ph/9504378} 
  M.~Spira, A.~Djouadi, D.~Graudenz and P.~M.~Zerwas,
  ``Higgs boson production at the LHC,''
  Nucl.\ Phys.\ B {\bf 453}, 17 (1995)
  [hep-ph/9504378].
  %%CITATION = HEP-PH/9504378;%%
  %850 citations counted in INSPIRE as of 04 Apr 2014


%\cite{hep-ph/9511344}
\bibitem{hep-ph/9511344} 
  A.~Djouadi, M.~Spira and P.~M.~Zerwas,
  ``QCD corrections to hadronic Higgs decays,''
  Z.\ Phys.\ C {\bf 70}, 427 (1996)
  [hep-ph/9511344].
  %%CITATION = HEP-PH/9511344;%%
  %161 citations counted in INSPIRE as of 04 Apr 2014


%\cite{hep-ph/9705337}
\bibitem{hep-ph/9705337} 
  M.~Spira,
  ``QCD effects in Higgs physics,''
  Fortsch.\ Phys.\  {\bf 46}, 203 (1998)
  [hep-ph/9705337].
  %%CITATION = HEP-PH/9705337;%%
  %402 citations counted in INSPIRE as of 04 Apr 2014


%\cite{hep-ph/0102227}
\bibitem{hep-ph/0102227} 
  S.~Catani, D.~de Florian and M.~Grazzini,
  ``Higgs production in hadron collisions: Soft and virtual QCD corrections at NNLO,''
  JHEP {\bf 0105}, 025 (2001)
  [hep-ph/0102227].
  %%CITATION = HEP-PH/0102227;%%
  %222 citations counted in INSPIRE as of 04 Apr 2014


%\cite{hep-ph/0201206}
\bibitem{hep-ph/0201206} 
  R.~V.~Harlander and W.~B.~Kilgore,
  ``Next-to-next-to-leading order Higgs production at hadron colliders,''
  Phys.\ Rev.\ Lett.\  {\bf 88}, 201801 (2002)
  [hep-ph/0201206].
  %%CITATION = HEP-PH/0201206;%%
  %754 citations counted in INSPIRE as of 04 Apr 2014


%\cite{hep-ph/0207004}
\bibitem{hep-ph/0207004} 
  C.~Anastasiou and K.~Melnikov,
  ``Higgs boson production at hadron colliders in NNLO QCD,''
  Nucl.\ Phys.\ B {\bf 646}, 220 (2002)
  [hep-ph/0207004].
  %%CITATION = HEP-PH/0207004;%%
  %683 citations counted in INSPIRE as of 04 Apr 2014


%\cite{hep-ph/0302135}
\bibitem{hep-ph/0302135} 
  V.~Ravindran, J.~Smith and W.~L.~van Neerven,
  ``NNLO corrections to the total cross-section for Higgs boson production in hadron hadron collisions,''
  Nucl.\ Phys.\ B {\bf 665}, 325 (2003)
  [hep-ph/0302135].
  %%CITATION = HEP-PH/0302135;%%
  %534 citations counted in INSPIRE as of 04 Apr 2014


%\cite{hep-ph/0509189}
\bibitem{hep-ph/0509189} 
  R.~Harlander and P.~Kant,
  ``Higgs production and decay: Analytic results at next-to-leading order QCD,''
  JHEP {\bf 0512}, 015 (2005)
  [hep-ph/0509189].
  %%CITATION = HEP-PH/0509189;%%
  %64 citations counted in INSPIRE as of 04 Apr 2014


%\cite{NUPHA.B510.61}
\bibitem{NUPHA.B510.61} 
  K.~G.~Chetyrkin, B.~A.~Kniehl and M.~Steinhauser,
  ``Decoupling relations to $\mathcal{O}\left(\als^3\right)$ and their connection to low-energy theorems,''
  Nucl.\ Phys.\ B {\bf 510}, 61 (1998)
  [hep-ph/9708255].
  %%CITATION = HEP-PH/9708255;%%
  %227 citations counted in INSPIRE as of 04 Apr 2014


%\cite{JHEPA.0601.051}
\bibitem{JHEPA.0601.051} 
  Y.~Schr\"{o}der and M.~Steinhauser,
  ``Four-loop decoupling relations for the strong coupling,''
  JHEP {\bf 0601}, 051 (2006)
  [hep-ph/0512058].
  %%CITATION = HEP-PH/0512058;%%
  %89 citations counted in INSPIRE as of 04 Apr 2014


%\cite{NUPHA.B744.121}
\bibitem{NUPHA.B744.121} 
  K.~G.~Chetyrkin, J.~H.~K\"{u}hn and C.~Sturm,
  ``QCD decoupling at four loops,''
  Nucl.\ Phys.\ B {\bf 744}, 121 (2006)
  [hep-ph/0512060].
  %%CITATION = HEP-PH/0512060;%%
  %86 citations counted in INSPIRE as of 04 Apr 2014


%\cite{PHLTA.B93.429}
\bibitem{PHLTA.B93.429} 
  O.~V.~Tarasov, A.~A.~Vladimirov and A.~Y.~Zharkov,
  ``The Gell-Mann-Low Function of QCD in the Three Loop Approximation,''
  Phys.\ Lett.\ B {\bf 93}, 429 (1980).
  %%CITATION = PHLTA,B93,429;%%
  %607 citations counted in INSPIRE as of 04 Apr 2014


%\cite{PHLTA.B303.334}
\bibitem{PHLTA.B303.334} 
  S.~A.~Larin and J.~A.~M.~Vermaseren,
  ``The Three loop QCD Beta function and anomalous dimensions,''
  Phys.\ Lett.\ B {\bf 303}, 334 (1993)
  [hep-ph/9302208].
  %%CITATION = HEP-PH/9302208;%%
  %315 citations counted in INSPIRE as of 04 Apr 2014


%\cite{PHLTA.B400.379}
\bibitem{PHLTA.B400.379} 
  T.~van Ritbergen, J.~A.~M.~Vermaseren and S.~A.~Larin,
  ``The Four loop beta function in quantum chromodynamics,''
  Phys.\ Lett.\ B {\bf 400}, 379 (1997)
  [hep-ph/9701390].
  %%CITATION = HEP-PH/9701390;%%
  %676 citations counted in INSPIRE as of 04 Apr 2014


%\cite{NUPHA.B710.485}
\bibitem{NUPHA.B710.485} 
  M.~Czakon,
  ``The Four-loop QCD beta-function and anomalous dimensions,''
  Nucl.\ Phys.\ B {\bf 710}, 485 (2005)
  [hep-ph/0411261].
  %%CITATION = HEP-PH/0411261;%%
  %189 citations counted in INSPIRE as of 04 Apr 2014

%\cite{JHEPA.1211.062}
\bibitem{JHEPA.1211.062} 
  C.~Anastasiou, S.~Buehler, C.~Duhr and F.~Herzog,
  ``NNLO phase space master integrals for two-to-one inclusive cross sections in dimensional regularization,''
  JHEP {\bf 1211}, 062 (2012)
  [arXiv:1208.3130 [hep-ph]].
  %%CITATION = ARXIV:1208.3130;%%
  %14 citations counted in INSPIRE as of 04 Apr 2014


%\cite{PHLTA.B721.244}
\bibitem{PHLTA.B721.244} 
  M.~H\"{o}schele, J.~Hoff, A.~Pak, M.~Steinhauser and T.~Ueda,
  ``Higgs boson production at the LHC: NNLO partonic cross sections through order $\epsilon$ and convolutions with splitting functions to N$^3$LO,''
  Phys.\ Lett.\ B {\bf 721}, 244 (2013)
  [arXiv:1211.6559 [hep-ph]].
  %%CITATION = ARXIV:1211.6559;%%
  %13 citations counted in INSPIRE as of 04 Apr 2014


%\cite{JHEPA.1310.096}
\bibitem{JHEPA.1310.096} 
  S.~Buehler and A.~Lazopoulos,
  ``Scale dependence and collinear subtraction terms for Higgs production in gluon fusion at N$^3$LO,''
  JHEP {\bf 1310}, 096 (2013)
  [arXiv:1306.2223 [hep-ph]].
  %%CITATION = ARXIV:1306.2223;%%
  %12 citations counted in INSPIRE as of 04 Apr 2014
%\cite{0902.3519}


\bibitem{0902.3519} 
  P.~A.~Baikov, K.~G.~Chetyrkin, A.~V.~Smirnov, V.~A.~Smirnov and M.~Steinhauser,
  ``Quark and gluon form factors to three loops,''
  Phys.\ Rev.\ Lett.\  {\bf 102}, 212002 (2009)
  [arXiv:0902.3519 [hep-ph]].
  %%CITATION = ARXIV:0902.3519;%%
  %60 citations counted in INSPIRE as of 04 Apr 2014


%\cite{1001.2887}
\bibitem{1001.2887} 
  R.~N.~Lee, A.~V.~Smirnov and V.~A.~Smirnov,
  ``Analytic Results for Massless Three-Loop Form Factors,''
  JHEP {\bf 1004}, 020 (2010)
  [arXiv:1001.2887 [hep-ph]].
  %%CITATION = ARXIV:1001.2887;%%
  %40 citations counted in INSPIRE as of 04 Apr 2014


%\cite{1004.3653}
\bibitem{1004.3653} 
  T.~Gehrmann, E.~W.~N.~Glover, T.~Huber, N.~Ikizlerli and C.~Studerus,
  ``Calculation of the quark and gluon form factors to three loops in QCD,''
  JHEP {\bf 1006}, 094 (2010)
  [arXiv:1004.3653 [hep-ph]].
  %%CITATION = ARXIV:1004.3653;%%
  %37 citations counted in INSPIRE as of 04 Apr 2014


%\cite{1010.4478}
\bibitem{1010.4478} 
  T.~Gehrmann, E.~W.~N.~Glover, T.~Huber, N.~Ikizlerli and C.~Studerus,
  ``The quark and gluon form factors to three loops in QCD through to $\mathcal{O}\left(\e^2\right)$,''
  JHEP {\bf 1011}, 102 (2010)
  [arXiv:1010.4478 [hep-ph]].
  %%CITATION = ARXIV:1010.4478;%%
  %14 citations counted in INSPIRE as of 04 Apr 2014

  
%\cite{1303.3590}
\bibitem{1303.3590} 
  R.~D.~Ball, M.~Bonvini, S.~Forte, S.~Marzani and G.~Ridolfi,
  ``Higgs production in gluon fusion beyond NNLO,''
  Nucl.\ Phys.\ B {\bf 874}, 746 (2013)
  [arXiv:1303.3590 [hep-ph]].
  %%CITATION = ARXIV:1303.3590;%%
  %25 citations counted in INSPIRE as of 17 Apr 2014cla
  
  
%\cite{1403.4616}
\bibitem{1403.4616} 
  C.~Anastasiou, C.~Duhr, F.~Dulat, E.~Furlan, T.~Gehrmann, F.~Herzog and B.~Mistlberger,
  ``Higgs boson gluon-fusion production at threshold in N$^3$LO QCD,''
  arXiv:1403.4616 [hep-ph].
  %%CITATION = ARXIV:1403.4616;%%
  %1 citations counted in INSPIRE as of 04 Apr 2014cla
  
  
%\cite{JHEPA.1202.056}
\bibitem{JHEPA.1202.056} 
  T.~Gehrmann, M.~Jaquier, E.~W.~N.~Glover and A.~Koukoutsakis,
  ``Two-Loop QCD Corrections to the Helicity Amplitudes for $H \to 3$ partons,''
  JHEP {\bf 1202}, 056 (2012)
  [arXiv:1112.3554 [hep-ph]].
  %%CITATION = ARXIV:1112.3554;%%
  %26 citations counted in INSPIRE as of 04 Apr 2014


%\cite{hep-ph/0405236}
\bibitem{hep-ph/0405236} 
  S.~D.~Badger and E.~W.~N.~Glover,
  ``Two loop splitting functions in QCD,''
  JHEP {\bf 0407}, 040 (2004)
  [hep-ph/0405236].
  %%CITATION = HEP-PH/0405236;%%
  %29 citations counted in INSPIRE as of 04 Apr 2014


%\cite{1309.4391}
\bibitem{1309.4391} 
  Y.~Li and H.~X.~Zhu,
  ``Single soft gluon emission at two loops,''
  JHEP {\bf 1311}, 080 (2013)
  [arXiv:1309.4391 [hep-ph]].
  %%CITATION = ARXIV:1309.4391;%%
  %3 citations counted in INSPIRE as of 04 Apr 2014


%\cite{1309.4393}
\bibitem{1309.4393} 
  C.~Duhr and T.~Gehrmann,
  ``The two-loop soft current in dimensional regularization,''
  Phys.\ Lett.\ B {\bf 727}, 452 (2013)
  [arXiv:1309.4393 [hep-ph]].
  %%CITATION = ARXIV:1309.4393;%%
  %4 citations counted in INSPIRE as of 04 Apr 2014


%\cite{JHEPA.1312.088}
\bibitem{JHEPA.1312.088} 
  C.~Anastasiou, C.~Duhr, F.~Dulat, F.~Herzog and B.~Mistlberger,
  ``Real-virtual contributions to the inclusive Higgs cross-section at N$^3$LO,''
  JHEP {\bf 1312}, 088 (2013)
  [arXiv:1311.1425 [hep-ph]].
  %%CITATION = ARXIV:1311.1425;%%
  %4 citations counted in INSPIRE as of 04 Apr 2014

  
%\cite{PHRVA.D60.116001}
\bibitem{PHRVA.D60.116001} 
  Z.~Bern, V.~Del Duca, W.~B.~Kilgore and C.~R.~Schmidt,
  ``The infrared behavior of one loop QCD amplitudes at next-to-next-to leading order,''
  Phys.\ Rev.\ D {\bf 60}, 116001 (1999)
  [hep-ph/9903516].
  %%CITATION = HEP-PH/9903516;%%
  %160 citations counted in INSPIRE as of 04 Apr 2014


%\cite{hep-ph/0007142}
\bibitem{hep-ph/0007142} 
  S.~Catani and M.~Grazzini,
  ``The soft gluon current at one loop order,''
  Nucl.\ Phys.\ B {\bf 591}, 435 (2000)
  [hep-ph/0007142].
  %%CITATION = HEP-PH/0007142;%%
  %107 citations counted in INSPIRE as of 04 Apr 2014
  

%\cite{Kilgore:2013gba}
\bibitem{1312.1296} 
  W.~B.~Kilgore,
  %``One-Loop Single-Real-Emission Contributions to $pp\to H + X$ at Next-to-Next-to-Next-to-Leading Order,''
  Phys.\ Rev.\ D {\bf 89}, 073008 (2014)
  [arXiv:1312.1296 [hep-ph]].
  %%CITATION = ARXIV:1312.1296;%%
  %11 citations counted in INSPIRE as of 03 Sep 2014
  

%\cite{1302.4379}
\bibitem{1302.4379} 
  C.~Anastasiou, C.~Duhr, F.~Dulat and B.~Mistlberger,
  ``Soft triple-real radiation for Higgs production at N$^3$LO,''
  JHEP {\bf 1307}, 003 (2013)
  [arXiv:1302.4379 [hep-ph]].
  %%CITATION = ARXIV:1302.4379;%%
  %20 citations counted in INSPIRE as of 04 Apr 2014


%\cite{hep-ph/0508265}
\bibitem{hep-ph/0508265} 
  S.~Moch and A.~Vogt,
  ``Higher-order soft corrections to lepton pair and Higgs boson production,''
  Phys.\ Lett.\ B {\bf 631}, 48 (2005)
  [hep-ph/0508265].
  %%CITATION = HEP-PH/0508265;%%
  %194 citations counted in INSPIRE as of 04 Apr 2014cla
  
  
%\cite{Shifman:1979eb}
\bibitem{Shifman:1979eb} 
  M.~A.~Shifman, A.~I.~Vainshtein, M.~B.~Voloshin and V.~I.~Zakharov,
  ``Low-Energy Theorems for Higgs Boson Couplings to Photons,''
  Sov.\ J.\ Nucl.\ Phys.\  {\bf 30}, 711 (1979)
  [Yad.\ Fiz.\  {\bf 30}, 1368 (1979)].
  %%CITATION = SJNCA,30,711;%%
  %473 citations counted in INSPIRE as of 04 Apr 2014


%\cite{hep-ph/0005275}
\bibitem{hep-ph/0005275} 
  C.~W.~Bauer, S.~Fleming and M.~E.~Luke,
  ``Summing Sudakov logarithms in $B \to X_s + \gamma$ in effective field theory,''
  Phys.\ Rev.\ D {\bf 63}, 014006 (2000)
  [hep-ph/0005275].
  %%CITATION = HEP-PH/0005275;%%
  %478 citations counted in INSPIRE as of 04 Apr 2014


%\cite{hep-ph/0011336}
\bibitem{hep-ph/0011336} 
  C.~W.~Bauer, S.~Fleming, D.~Pirjol and I.~W.~Stewart,
  ``An Effective field theory for collinear and soft gluons: Heavy to light decays,''
  Phys.\ Rev.\ D {\bf 63}, 114020 (2001)
  [hep-ph/0011336].
  %%CITATION = HEP-PH/0011336;%%
  %757 citations counted in INSPIRE as of 04 Apr 2014


%\cite{hep-ph/0109045}
\bibitem{hep-ph/0109045} 
  C.~W.~Bauer, D.~Pirjol and I.~W.~Stewart,
  ``Soft collinear factorization in effective field theory,''
  Phys.\ Rev.\ D {\bf 65}, 054022 (2002)
  [hep-ph/0109045].
  %%CITATION = HEP-PH/0109045;%%
  %632 citations counted in INSPIRE as of 04 Apr 2014


%\cite{hep-ph/0206152}
\bibitem{hep-ph/0206152} 
  M.~Beneke, A.~P.~Chapovsky, M.~Diehl and T.~Feldmann,
  ``Soft collinear effective theory and heavy to light currents beyond leading power,''
  Nucl.\ Phys.\ B {\bf 643}, 431 (2002)
  [hep-ph/0206152].
  %%CITATION = HEP-PH/0206152;%%
  %334 citations counted in INSPIRE as of 04 Apr 2014
  
  
%\cite{hep-ph/0605068}
\bibitem{hep-ph/0605068} 
  A.~Idilbi, X.~Ji and F.~Yuan,
  ``Resummation of threshold logarithms in effective field theory for DIS, Drell-Yan and Higgs production,''
  Nucl.\ Phys.\ B {\bf 753}, 42 (2006)
  [hep-ph/0605068].
  %%CITATION = HEP-PH/0605068;%%
  %57 citations counted in INSPIRE as of 04 Apr 2014


%\cite{0710.0680}
\bibitem{0710.0680} 
  T.~Becher, M.~Neubert and G.~Xu,
  ``Dynamical Threshold Enhancement and Resummation in Drell-Yan Production,''
  JHEP {\bf 0807}, 030 (2008)
  [arXiv:0710.0680 [hep-ph]].
  %%CITATION = ARXIV:0710.0680;%%
  %113 citations counted in INSPIRE as of 04 Apr 2014


%\cite{0808.3008}
\bibitem{0808.3008} 
  V.~Ahrens, T.~Becher, M.~Neubert and L.~L.~Yang,
  ``Origin of the Large Perturbative Corrections to Higgs Production at Hadron Colliders,''
  Phys.\ Rev.\ D {\bf 79}, 033013 (2009)
  [arXiv:0808.3008 [hep-ph]].
  %%CITATION = ARXIV:0808.3008;%%
  %74 citations counted in INSPIRE as of 04 Apr 2014


\bibitem{hep-ph/0412110} 
  J.~Chay, C.~Kim, Y.~G.~Kim and J.~-P.~Lee,
  ``Soft Wilson lines in soft-collinear effective theory,''
  Phys.\ Rev.\ D {\bf 71}, 056001 (2005)
  [hep-ph/0412110].
  %%CITATION = HEP-PH/0412110;%%
  %25 citations counted in INSPIRE as of 29 Mar 2014
  
  
%\cite{hep-ph/9808389}
\bibitem{hep-ph/9808389} 
  A.~V.~Belitsky,
  ``Two loop renormalization of Wilson loop for Drell-Yan production,''
  Phys.\ Lett.\ B {\bf 442}, 307 (1998)
  [hep-ph/9808389].
  %%CITATION = HEP-PH/9808389;%%
  %21 citations counted in INSPIRE as of 04 Apr 2014


  %\cite{Gatheral:1983cz}
\bibitem{PHLTA.B133.90} 
  J.~G.~M.~Gatheral,
  ``Exponentiation of Eikonal Cross-sections in Nonabelian Gauge Theories,''
  Phys.\ Lett.\ B {\bf 133}, 90 (1983).
  %%CITATION = PHLTA,B133,90;%%
  %181 citations counted in INSPIRE as of 15 Apr 2014
  
  
%\cite{Frenkel:1984pz}
\bibitem{NUPHA.B246.231} 
  J.~Frenkel and J.~C.~Taylor,
  ``Nonabelian Eikonal Exponentiation,''
  Nucl.\ Phys.\ B {\bf 246}, 231 (1984).
  %%CITATION = NUPHA,B246,231;%%
  %176 citations counted in INSPIRE as of 15 Apr 2014

%\cite{Nogueira:1991ex}
\bibitem{Nogueira:1991ex} 
  P.~Nogueira,
  ``Automatic Feynman graph generation,''
  J.\ Comput.\ Phys.\  {\bf 105}, 279 (1993).
  %%CITATION = JCTPA,105,279;%%
  %470 citations counted in INSPIRE as of 04 Apr 2014


%\cite{math-ph/0010025}
\bibitem{math-ph/0010025} 
  J.~A.~M.~Vermaseren,
  ``New features of FORM,''
  math-ph/0010025.
  %%CITATION = MATH-PH/0010025;%%
  %966 citations counted in INSPIRE as of 04 Apr 2014

  
\bibitem{PHLTA.B100.65} 
  F.~V.~Tkachov,
  ``A Theorem on Analytical Calculability of Four Loop Renormalization Group Functions,''
  Phys.\ Lett.\ B {\bf 100}, 65 (1981).
  %%CITATION = PHLTA,B100,65;%%
  %587 citations counted in INSPIRE as of 01 Apr 2014
  
  
%\cite{NUPHA.B192.159}
\bibitem{NUPHA.B192.159} 
  K.~G.~Chetyrkin and F.~V.~Tkachov,
  ``Integration by Parts: The Algorithm to Calculate beta Functions in 4 Loops,''
  Nucl.\ Phys.\ B {\bf 192}, 159 (1981).
  %%CITATION = NUPHA,B192,159;%%
  %892 citations counted in INSPIRE as of 07 Mar 2014
  
  
%\cite{Laporta:2001dd}
\bibitem{hep-ph/0102033} 
  S.~Laporta,
  ``High precision calculation of multiloop Feynman integrals by difference equations,''
  Int.\ J.\ Mod.\ Phys.\ A {\bf 15}, 5087 (2000)
  [hep-ph/0102033].
  %%CITATION = HEP-PH/0102033;%%
  %433 citations counted in INSPIRE as of 09 Apr 2014
  
  
%\cite{1201.4330}
\bibitem{1201.4330} 
  A.~von Manteuffel and C.~Studerus,
  ``Reduze 2 - Distributed Feynman Integral Reduction,''
  arXiv:1201.4330 [hep-ph].
  %%CITATION = ARXIV:1201.4330;%%
  %32 citations counted in INSPIRE as of 04 Apr 2014
  
  
%\cite{hep-ph/0306192}
\bibitem{hep-ph/0306192} 
  C.~Anastasiou, L.~J.~Dixon, K.~Melnikov and F.~Petriello,
  ``Dilepton rapidity distribution in the Drell-Yan process at NNLO in QCD,''
  Phys.\ Rev.\ Lett.\  {\bf 91}, 182002 (2003)
  [hep-ph/0306192].
  %%CITATION = HEP-PH/0306192;%%
  %132 citations counted in INSPIRE as of 04 Apr 2014


%\cite{hep-ph/0312266}
\bibitem{hep-ph/0312266} 
  C.~Anastasiou, L.~J.~Dixon, K.~Melnikov and F.~Petriello,
  ``High precision QCD at hadron colliders: Electroweak gauge boson rapidity distributions at NNLO,''
  Phys.\ Rev.\ D {\bf 69}, 094008 (2004)
  [hep-ph/0312266].
  %%CITATION = HEP-PH/0312266;%%
  %438 citations counted in INSPIRE as of 04 Apr 2014
  

\bibitem{Exton}
  H.~Exton,
  ``Handbook of Hypergeometric Integrals,"
   Ellis Hornwood Limited, Chichester, (1978).

   
%\cite{hep-ph/0004013}
\bibitem{hep-ph/0004013} 
  T.~Binoth and G.~Heinrich,
  ``An automatized algorithm to compute infrared divergent multiloop integrals,''
  Nucl.\ Phys.\ B {\bf 585}, 741 (2000)
  [hep-ph/0004013].
  %%CITATION = HEP-PH/0004013;%%
  %238 citations counted in INSPIRE as of 04 Apr 2014
  

%\cite{Bogner:2007cr}
\bibitem{0709.4092} 
  C.~Bogner and S.~Weinzierl,
  ``Resolution of singularities for multi-loop integrals,''
  Comput.\ Phys.\ Commun.\  {\bf 178}, 596 (2008)
  [arXiv:0709.4092 [hep-ph]].
  %%CITATION = ARXIV:0709.4092;%%
  %59 citations counted in INSPIRE as of 11 Apr 2014
  

%\cite{Smirnov:2013eza}
\bibitem{1312.3186} 
  A.~V.~Smirnov,
  %``FIESTA 3: cluster-parallelizable multiloop numerical calculations in physical regions,''
  Comput.\ Phys.\ Commun.\  {\bf 185}, 2090 (2014)
  [arXiv:1312.3186 [hep-ph]].
  %%CITATION = ARXIV:1312.3186;%%
  %7 citations counted in INSPIRE as of 03 Sep 2014
  

%\cite{Kelley:2011ng}
\bibitem{1105.3676} 
  R.~Kelley, M.~D.~Schwartz, R.~M.~Schabinger and H.~X.~Zhu,
  ``The two-loop hemisphere soft function,''
  Phys.\ Rev.\ D {\bf 84}, 045022 (2011)
  [arXiv:1105.3676 [hep-ph]].
  %%CITATION = ARXIV:1105.3676;%%
  %33 citations counted in INSPIRE as of 11 Apr 2014
  

%\cite{Kelley:2011aa}
\bibitem{1112.3343} 
  R.~Kelley, M.~D.~Schwartz, R.~M.~Schabinger and H.~X.~Zhu,
  ``Jet Mass with a Jet Veto at Two Loops and the Universality of Non-Global Structure,''
  Phys.\ Rev.\ D {\bf 86}, 054017 (2012)
  [arXiv:1112.3343 [hep-ph]].
  %%CITATION = ARXIV:1112.3343;%%
  %13 citations counted in INSPIRE as of 11 Apr 2014
  
  
%\cite{vonManteuffel:2013vja}
\bibitem{1309.3560} 
  A.~von Manteuffel, R.~M.~Schabinger and H.~X.~Zhu,
  ``The Complete Two-Loop Integrated Jet Thrust Distribution In Soft-Collinear Effective Theory,''
  JHEP {\bf 1403}, 139 (2014)
  [arXiv:1309.3560 [hep-ph]].
  %%CITATION = ARXIV:1309.3560;%%
  %3 citations counted in INSPIRE as of 11 Apr 2014


%\cite{hep-ph/0507094}
\bibitem{hep-ph/0507094} 
  T.~Huber and D.~Maitre,
  ``HypExp: A Mathematica package for expanding hypergeometric functions around integer-valued parameters,''
  Comput.\ Phys.\ Commun.\  {\bf 175}, 122 (2006)
  [hep-ph/0507094].
  %%CITATION = HEP-PH/0507094;%%
  %105 citations counted in INSPIRE as of 04 Apr 2014
 
 
\bibitem{0708.2443} 
  T.~Huber and D.~Maitre,
  ``HypExp 2, Expanding Hypergeometric Functions about Half-Integer Parameters,''
  Comput.\ Phys.\ Commun.\  {\bf 178}, 755 (2008)
  [arXiv:0708.2443 [hep-ph]].
  %%CITATION = ARXIV:0708.2443;%%
  %79 citations counted in INSPIRE as of 09 Apr 2014
  

%\cite{Bern:2004cz}
\bibitem{hep-ph/0404293} 
  Z.~Bern, L.~J.~Dixon and D.~A.~Kosower,
  ``Two-loop $g \to gg$ splitting amplitudes in QCD,''
  JHEP {\bf 0408}, 012 (2004)
  [hep-ph/0404293].
  %%CITATION = HEP-PH/0404293;%%
  %128 citations counted in INSPIRE as of 19 Apr 2014cla
  

%\cite{1304.1806}
\bibitem{1304.1806} 
  J.~M.~Henn,
  ``Multiloop integrals in dimensional regularization made simple,''
  Phys.\ Rev.\ Lett.\  {\bf 110}, no. 25, 251601 (2013)
  [arXiv:1304.1806 [hep-th]].
  %%CITATION = ARXIV:1304.1806;%%
  %17 citations counted in INSPIRE as of 07 Mar 2014
  

%\cite{Lee:2010ik}
\bibitem{1010.1334} 
  R.~N.~Lee and V.~A.~Smirnov,
  ``Analytic Epsilon Expansions of Master Integrals Corresponding to Massless Three-Loop Form Factors and Three-Loop $\mathrm{g}-2$ up to Four-Loop Transcendentality Weight,''
  JHEP {\bf 1102}, 102 (2011)
  [arXiv:1010.1334 [hep-ph]].
  %%CITATION = ARXIV:1010.1334;%%
  %15 citations counted in INSPIRE as of 11 Apr 2014
  
  
%\cite{Matsuura:1988sm}
\bibitem{NUPHA.B319.570} 
  T.~Matsuura, S.~C.~van der Marck and W.~L.~van Neerven,
  ``The Calculation of the Second Order Soft and Virtual Contributions to the Drell-Yan Cross-Section,''
  Nucl.\ Phys.\ B {\bf 319}, 570 (1989).
  %%CITATION = NUPHA,B319,570;%%
  %194 citations counted in INSPIRE as of 11 Apr 2014
  
  
%\cite{Harlander:2000mg}
\bibitem{hep-ph/0007289} 
  R.~V.~Harlander,
  ``Virtual corrections to $g g \to H$ to two loops in the heavy top limit,''
  Phys.\ Lett.\ B {\bf 492}, 74 (2000)
  [hep-ph/0007289].
  %%CITATION = HEP-PH/0007289;%%
  %160 citations counted in INSPIRE as of 11 Apr 2014
  
 
%\cite{Laenen:2005uz}
\bibitem{hep-ph/0508284} 
  E.~Laenen and L.~Magnea,
  ``Threshold resummation for electroweak annihilation from DIS data,''
  Phys.\ Lett.\ B {\bf 632}, 270 (2006)
  [hep-ph/0508284].
  %%CITATION = HEP-PH/0508284;%%
  %97 citations counted in INSPIRE as of 11 Apr 2014
  
  
%\cite{Ravindran:2006cg}
\bibitem{hep-ph/0603041} 
  V.~Ravindran,
  ``Higher-order threshold effects to inclusive processes in QCD,''
  Nucl.\ Phys.\ B {\bf 752}, 173 (2006)
  [hep-ph/0603041].
  %%CITATION = HEP-PH/0603041;%%
  %67 citations counted in INSPIRE as of 11 Apr 2014
  
  
%\cite{Ahrens:2008nc}
\bibitem{0809.4283} 
  V.~Ahrens, T.~Becher, M.~Neubert and L.~L.~Yang,
  ``Renormalization-Group Improved Prediction for Higgs Production at Hadron Colliders,''
  Eur.\ Phys.\ J.\ C {\bf 62}, 333 (2009)
  [arXiv:0809.4283 [hep-ph]].
  %%CITATION = ARXIV:0809.4283;%%
  %104 citations counted in INSPIRE as of 11 Apr 2014
  
  
%\cite{Ahmed:2014cla}
\bibitem{1404.0366} 
  T.~Ahmed, M.~Mahakhud, N.~Rana and V.~Ravindran,
  ``Drell-Yan production at threshold in N$^3$LO QCD,''
  arXiv:1404.0366 [hep-ph].
  %%CITATION = ARXIV:1404.0366;%%
  

%\cite{Binosi:2003yf}
\bibitem{hep-ph/0309015} 
  D.~Binosi and L.~Theussl,
  ``JaxoDraw: A Graphical user interface for drawing Feynman diagrams,''
  Comput.\ Phys.\ Commun.\  {\bf 161}, 76 (2004)
  [hep-ph/0309015].
  %%CITATION = HEP-PH/0309015;%%
  %297 citations counted in INSPIRE as of 19 Apr 2014cla
  
  
%\cite{Vermaseren:1994je}
\bibitem{CPHCB.83.45} 
  J.~A.~M.~Vermaseren,
  ``Axodraw,''
  Comput.\ Phys.\ Commun.\  {\bf 83}, 45 (1994).
  %%CITATION = CPHCB,83,45;%%
  %127 citations counted in INSPIRE as of 19 Apr 2014
\end{thebibliography}
\end{document}